\documentclass{ws-ijmpa}
\def\x{\times}
\def\G{{\cal G}}
\def\r2{\sqrt{2}}
\def\rs{\sqrt{6}}
\def\rt{\sqrt{3}}
\def\be{\begin{eqnarray}}
\def\ee{\end{eqnarray}}
\def\bk#1{\langle#1\rangle}
\def\l{\lambda}

\begin{document}


%
\catchline{}{}{}{}{}
%

\title{HORIZONTAL SYMMETRY}

\author{C.S. Lam}

\address{Physics Department, McGill University,
Montreal, Canada\\
{\it and}\\
Department of Physics and Astronomy, University of British Columbia,
Vancouver, Canada\\
Lam@physics.mcgill.ca}

\maketitle


\begin{abstract}
The relation between fermion mixing and horizontal symmetry is discussed.

\end{abstract}


\section{Introduction}	
Symmetry considerations have helped to create the Standard Model,
can it also help us understand the mystery of three fermion generations? 
We shall see that it might for
fermion mixing but not for their masses. Further details and references can be found in Ref.~[1].

To simplify matters, I shall assume at the outset that all right-handed and heavy 
fermions have been integrated out so we are only left with the left-handed fermions
to deal with. The effective $3\x 3$ mass matrices are then $M_uM_u^\dagger, M_dM_d^\dagger, M_eM_e^\dagger$, and $M_\nu$,
respectively for the up and the  down quarks, the charged leptons and
the Majorana neutrinos.
I shall discuss the quark sector explicitly, but the 
conclusion thus reached is also valid in the lepton sector.

The mixing matrix for the quarks is $U=W^\dagger V$, where $W,V$ are the unitary matrices which
render $W^\dagger M_uM_u^\dagger W$ and $V^\dagger M_dM_d^\dagger V$ diagonal. Equivalently,
the three columns of $W$ are the three eigenvectors of $M_uM_u^\dagger$, and the three columns of $V$ are the 
three eigenvectors of $M_dM_d^\dagger$.
Since the quark masses are all
different, these eigenvectors are unique up to arbitrary phases.

A horizontal symmetry is defined by a unitary matrix $G$ which commutes with the mass matrices $M_uM_u^\dagger$
and $M_dM_d^\dagger$. As such, it must have the same eigenvectors as $M_uM_u^\dagger$, and 
also the same eigenvectors as $M_dM_d^\dagger$. If the eigenvalues of $G$ are non-degenerate, then these
two sets of eigenvectors must be the same. We can then set $W=V$, which results in $U=1$ and no mixing.
Even if an eigenvalue of $G$ is doubly degenerate, the eigenvector for its non-degenerate eigenvalue must be
common to $G, M_uM_u^\dagger$, and $M_dM_d^\dagger$, hence one of the three quarks does not mix. Neither agrees
with Nature so it is not possible to have a horizontal symmetry common to the up and the down quarks.
The best we can do is to have a horizontal symmetry $F$ for the up quarks, and a different horizontal symmetry $G$
for the down quarks. Thus whatever horizontal group  
that may be present at high energies must be broken down at present energies to these
{\it residual symmetries}
$F$ and $G$. The original horizontal symmetry group, which 
we shall assume to be a finite group, must contain $F$ and $G$ and hence the finite group $\G=\{F,G\}$ generated by them.

Our task here is to find out the connection
 between the mixing matrix $U$ and the horizontal symmetry group $\G$. In the next
section we discuss what are the possible $\G$'s for a given $U$. In the section after we discuss what are the
possible $U$'s for a given $\G$, how to construct the effective mass terms invariant under $\G$, and 
how to break them spontaneous to obtain the mixing patterns. Realistic neutrino and quark mixings are used
to illustrate these relations.

\section{From $U$ to $\G$} 
Suppose $U$ is given, say by experiments, then its
horizontal symmetry group $\G=\{F,G\}$ is determined by
the residual symmetries $F$ and $G$. To find them, it is convenient to work
in the basis where $M_uM_u^\dagger$ is diagonal. In that base $W=1$ and $U=V$, so the eigenvectors of $G$ are given by the
columns of $U$. Since $\G$ is a finite group, there must be an integer $k$ so that $G^k=1$, hence the eigenvalues of
$G$ are the $k$th roots of unity. As to $F$, it must be diagonal to be able to commute with $M_uM_u^\dagger$, and its
entries must all be the $n$th root of unity if $F^n=1$. Thus there is a series of compatible $F$
and $G$ for each $U$, labeled by the integers $k$ and $n$ as well as their eigenvalues.
We shall reject those $(F,G)$ pairs 
which do not generate a finite group.

The eigenvalues for $F$ and $G$ can each be chosen to be degenerate or non-degenerate. The former means that two of the three
eigenvalues are the same, and the latter means that all three are different. These two situations
are qualitatively distinct
when we try to recover $U$ from a horizontal group $\G$, a topic we shall discuss in the next section. Since $U$ is the diagonalization
matrix for $M_dM_d^\dagger$ when $M_uM_u^\dagger$ is diagonal, we need to know $F$ and $G$ in that basis. This can be achieved
if $F$ is non-degenerate, for then in the representation when $F$ is diagonal, $M_uM_u^\dagger$ is automatically diagonal since it commutes
with $F$. This is no longer the case if $F$ is degenerate, so in order to be able to recover $U$ from $\G$, we shall
require $F$ to be non-degenerate. What about $G$? If it is non-degenerate, then its eigenvectors
assembled together constitute $U$. If it is degenerate, as it will be forced to be so for leptons,  then
only the eigenvector associated with its non-degenerate eigenvalue is fixed, 
so we can recover only one column of $U$. In order to
obtain the whole $U$, $\G$ needs to contain another residual symmetry 
$G'$ which commutes with $G$, but with its non-degenerate eigenvalue in a different
position than that of $G$. Then the non-degenerate eigenvector of $G'$ yields a second column of $U$. Since $U$ is unitary,
its third column must be orthogonal to these two and is automatically determined once these two are given. In this way $U$
can be completely recovered.  By the way, when I say $U$ is determined from $\G$, 
I always mean that its column vectors are known, but
the ordering of them in $U$ is never certain. That must be determined by experiments from the physics context.

All of these work for the leptons as well, though with one additional constraint already hinted above. 
The Majorana neutrino mass matrix is symmetric, as such it  
is diagonalized by a unitary matrix $V$, but now to render $V^TM_\nu V$ 
but not $V^\dagger M_\nu V$ diagonal. This forces  $G$ to have eigenvalues $\pm 1$. We shall 
choose the sign of $G$ so that it has a single eigenvalue $+1$ and two eigenvalues  $-1$. For a given $U$ with columns $v_i$,
three such $G$'s exists and they are $G_i=-1+2v_iv_i^\dagger$. It can be verified that these three mutually commute, and 
the product of two of them is equal to the third, so any two can be chosen to be the $G$ and $G'$ discussed at the end of the
last paragraph. If $\G$ contains only one $G_i$, then that horizontal symmetry group can only predict one column of $U$, leaving
the other two determined only by unitarity.

Let me now illustrate these general discussions with two examples, one for neutrino mixing, and one for quark mixing.

\subsection{Neutrino Mixing}
Within experimental errors, neutrino mixing can be described by the tribimaximal PMNS mixing matrix\cite{HPS}
\be
U={1\over\rs}\begin{pmatrix}2&\r2&0\cr -1&\r2&\rt\cr -1&\r2&-\rt\cr\end{pmatrix},\ee
from which we can work out the three $G_i$'s. With $F^3=1$, the horizontal groups $\G_3^i=\{F,G_i\}$ are
\be
i=1: S_4, \Delta(216); \quad i=2: A_4;\quad i=3: S_3, \Delta'(54),\ee
where $S_n$ is the  symmetric (permutation) group of $n$ objects and $A_n$ is the alternating
(even permutation) group of $n$ objects. $\Delta(216)$ is known as the Hessian group and is a finite
subgroup of $SU(3)$ with order 216, and $\Delta'(54)$ is one of its subgroups with order 54. The group generated
by $\{F,G,G'\}$ is identical to the group for $i=1$.

\subsection{Quark Mixing}
We discuss here the mixing of two quarks; the mixing of three quarks is being investigated. For two quarks, the Cabibbo mixing
matrix is
\be
U=\begin{pmatrix}c&s\cr -s&c\cr\end{pmatrix},\ee
where $c=\cos\theta_c$, $s=\sin\theta_c$, and $\theta_c$ is the Cabibbo angle. We may choose $F={\rm diag}(1,-1)$
which is non-degenerate, and for two generations there is only one $G$. It can be worked out from $U$ to be 
\be
G=\begin{pmatrix}c_2&-s_2\cr -s_2&-c_2\cr\end{pmatrix},\quad\Rightarrow 
(GF)^m=\begin{pmatrix}c_{2m}&s_{2m}\cr -s_{2m}&c_{2m}\cr\end{pmatrix},\ee
where $c_n=\cos(n\theta_c)$ and $s_n=\sin(n\theta_c)$. The group $\G=\{F,G\}$ is a finite group if $(GF)^m=1$ for some integer $m$; the group
so generated is the dihedral group $D_{m}$. According to (4), this calls for $\theta_c=\pi/2m$. 
For $m=7$, the value for $\l=\sin\theta_c$
is 0.2225, which differs  from the experimental value by something much less than $\l^2/2$, an error which we might expect to meet
by neglecting the mixing with the third quark. 

\section{From $\G$ to $U$}
For a finite group to be a horizontal symmetry group, clearly it must contain a three-dimensional representation.
If the representation is reducible to three one-dimensional representations, then there is no mixing. If it is reducible to a
one- and a two-dimensional irreducible representations, then the mixing matrix must have at least one zero\cite{LAM}, 
like that shown in eq.~(1). So, for a generic mixing matrix, we must consider groups $\G$ with a three-dimensional irreducible 
representation. These groups are all finite subgroups of $SU(2)$ or $SU(3)$, and they are all known. Given one of these groups,
any pair of group elements $(F,G)$ such that $F$ is non-degenerate is eligible to be the residual operators. For neutrino mixing, we
also require $G$ to have order two. Once this pair is chosen, we can compute the mixing matrix in the basis where $F$
is diagonal. If $G$ is non-degenerate, then $U$ is completely known. Otherwise, one column per $G$ can be so extracted.

We also need to know how the full horizontal symmetry $\G$ operates in the symmetric phase, and how it is spontaneously 
broken to the residual symmetries $(F,G)$  to obtain the desired 
mixing matrix $U$. Remembering that all the right-handed and heavy fermions have already been integrated out,
the charged-lepton ($e$) mass terms in the symmetric phase are therefore of the form $C_{ija}(e^\dagger)_ie_j(\phi_e)_a$,
and the Majorana neutrino ($\nu$) mass terms are of the form $C'_{ija}(\nu^T)_i\nu_j(\phi_\nu)_a$, where 
$C_{ija}$ and $C'_{ija}$ are the Clebsch-Gordan coefficients to ensure a $\G$-invariant coupling, and the Higgs fields
$\phi_e,\phi_\nu$ may be simple or composite, and may be in any irreducible or reducible representations.

The breakdown to residual symmetries is achieved by introducing
suitable vacuum expectation values to the Higgs fields, whence the charged-lepton
mass matrix becomes $(M_eM_e^\dagger)_{ij}=C_{ija}\bk{(\phi_e)_a}$, and the neutrino mass matrix becomes $(M_\nu)_{ij}=C'_{ija}\bk{(\phi_\nu)_a}$.
In order for $F$ and $G$ to be residual symmetries of the mass matrices, we must require
\be
F''\bk{\phi_e}=\bk{\phi_e},\qquad G''\bk{\phi_\nu}=\bk{\phi_\nu},\ee
where $F''$ and $G''$ are $F$ and $G$ in the representation of $\bk{\phi_e}$ and $\bk{\phi_\nu}$, respectively.

This condition is illustrated below in the case of neutrino mixing with the horizontal
symmetry group $\G=A_4$. From (2), we can see that this
group corresponds to $i=2$, so the mixing matrix $U$ is expected to have the trimaximal mixing exhibited in the
second column of (1). Indeed it does, if we go through the calculation using (5). 

\subsection{From $A_4$ to trimaximal mixing}
$A_4$ has three 1-dimensional and one 3-dimensional irreducible representations, labeled as ${\bf 1,1',1'',3}$.
We follow Ref.~[3] by taking $\phi_e$ to belong to {\bf 3}, and six $\phi_\nu$ fields to belong to ${\bf 1,1',1'',3}$.
The 3-dimensional representation from (1) and (2) are
\be
F=\begin{pmatrix}1&0&0\cr 0&\omega&0\cr 0&0&\omega^2\cr\end{pmatrix},\qquad 
G=G_2={1\over 3}\begin{pmatrix}-1&2&2\cr 2&-1&2\cr 2&2&-1\cr\end{pmatrix},\ee
where $\omega$ is a non-trivial cube root of unity. To satisfy (5) for the triplet ({\bf 3}) Higgs,
we need $\bk{\phi_e}$ to be proportional to $(1,0,0)^T$, and $\bk{\phi_\nu^{(3)}}$ to be proportional to 
$(1,1,1)^T$. The singlet representations $({\bf 1,1', 1''})$ for $G_2$ turns out to be 1 for all three of them,
hence (5) is automatically satisfied for $\bk{\phi_\nu^{(1,1',1'')}}$.

\section{Conclusion}
There is a connection between horizontal symmetry and fermion mixing, though none with fermion masses. The relation
between horizontal symmetry and mixing is not simple: it is neither one to one, nor can the symmetry be unbroken.
Nevertheless, from a mixing matrix $U$ one can determine in principle all the finite horizontal groups $\G$
that can lead to it. 
Conversely, 
from a given finite horizontal group $\G$ one can work out all the possible mixing patterns $U$.
After the right-handed and heavy fermions are integrated out, 
we can construct the general mass terms invariant under $\G$,   and work out how the full symmetry
can be broken down to the residual symmetries of a particular mixing pattern. This general theory is
illustrated with realistic neutrino and quark mixings.


\begin{thebibliography}{0}    
\bibitem{LAM} C.S. Lam, arxiv: 0708.3665, to appear in Phys. Lett.
\bibitem{HPS} P.F. Harrison, D.H. Perkins, and W.G. Scott, Phys. Lett. B458 (1999) 79.
\bibitem{MA} E. Ma, Phys.~Rev. D70 (2004) 031901; G. Altarelli and F. Feruglio, Nucl.~Phys. B741 (2006) 215. The
representation for $F$ and $G$ used here are the those in the second reference.
\end{thebibliography}
\end{document}